# On optimization of Paganin's method for propagation-based
# X-ray phase-contrast imaging and tomography

Timur E. Gureyev [a)], David M. Paganin [b)], Ashkan Pakzad [a)] and Harry M. Quiney [a)]

[a)] School of Physics, University of Melbourne, Parkville, Victoria, 3010, Australia

[b)] School of Physics and Astronomy, Monash University, Clayton, Victoria, 3800, Australia

Correspondence email: timur.gureyev@unimelb.edu.au

Funding information: National Health and Medical Research Council, Australia (grant No. APP2011204).

## Abstract

Paganin's method for image reconstruction in propagation-based phase-contrast X-ray imaging and tomography has enjoyed broad acceptance in recent years, with over one thousand publications citing its use. The present paper discusses approaches to optimization of the method with respect to simple image quality metrics, such as signal-to-noise ratio and spatial resolution, as well as a reference-based metric corresponding to the relative mean squared difference between the reconstructed image and the "ground truth" image that would be obtained in a setup with perfect spatial resolution and no noise. The problem of optimization of the intrinsic regularization parameter of Paganin's method with respect to spatial resolution in the reconstructed image is studied in detail. It is also demonstrated that a combination of Paganin's method with a Tikhonov-regularized deconvolution of the point-spread function of the imaging system can provide significantly higher image quality compared to the standard version of the method. Analytical expressions for some relevant image quality metrics are obtained and compared with results of numerical simulations. Advantages and shortcomings of optimization approaches using a number of different image quality metrics are discussed. The results of this study are expected to be useful in practical X-ray imaging and training of deep machine learning models for image denoising and segmentation.

## 1. Introduction

Propagation-based X-ray phase-contrast imaging (PBI) and computed tomography (PB-CT) are popular techniques for imaging samples that exhibit low X-ray attenuation contrast (Snigirev *et* al., 1995; Wilkins *et* al., 1996; Cloetens *et al.*, 1996; Paganin, 2006). These methods are particularly useful in imaging biological samples, where they can considerably improve the contrast and signal-to-noise (SNR) in images of soft biological tissues in comparison with conventional absorption-based X-ray imaging at the same radiation dose and spatial resolution (Wilkins *et al.*, 2014; Endrizzi, 2018; Momose, 2020; Quenot *et al.*, 2022). In recent years, PBI techniques have been under active development for medical X-ray imaging applications, where they offer the promise of reducing radiation dose and improving image quality compared to existing medical X-ray imaging methods (Taba *et al.*, 2020; Longo *et al.*, 2024; Donato *et al.*, 2025; Gureyev *et al.*, 2025a; Pakzad *et al.*, 2025). In this context, it is important to investigate hardware setups and software algorithms that optimize PBI and PB-CT image acquisition and processing techniques and maximize their diagnostic performance.

In a recent paper (Gureyev *et al.*, 2025b), we derived analytical expressions for optimizing PBI and PB-CT imaging setups – including the X-ray source size and energy, the detector resolution, the source-to-sample and sample-to-detector distances – with the goal of maximizing the quality of collected images at a fixed radiation dose. The image quality was evaluated in terms of contrast, SNR, spatial resolution and the recently introduced biomedical X-ray imaging quality characteristic (Gureyev *et al.*, 2025a) that combines the contrast-to-noise, spatial resolution and radiation dose into a single metric. These results were based on application of the standard form of the Homogeneous Transport of Intensity equation (TIE-Hom) formalism for image reconstruction, which is often referred to as "Paganin's method" (Paganin *et al.*, 2002). This method was used for reconstruction of equivalent object-plane images from PBI images collected at different sample-to-detector distances. In contrast, the present paper is focused on the optimization of Paganin's method as an image processing technique. Initially, the optimization is considered in this paper with respect to the intrinsic regularization parameter of Paganin's method. This parameter is theoretically determined by the X-ray wavelength, the effective defocus distance and the ratio of the real decrement to the imaginary part of the complex X-ray refractive index of the imaged sample. Optimization of the TIE-Hom regularization parameter was previously considered by Beltran *et al.* (2018), where it was based in part on the "deblur-by-defocus" method proposed by Gureyev *et al*. (2004). These approaches exploit the fact that the point-spread function (PSF) of a PBI system effectively performs partial "phase retrieval" in hardware by blurring the propagated images in a way that directly counteracts the image sharpening that occurs as a result of coherent free-space propagation (Gureyev *et al*., 2004; Beltran *et al.*, 2018). The subsequent TIE-Hom retrieval of collected images in software consequently needs to be performed with a reduced value of the regularization parameter in order to achieve the optimal result. After investigating this problem, we then proceed with a more general optimization of Paganin's method that can be achieved by combining the TIE-Hom retrieval with an explicit regularized deconvolution of the system PSF. We show that this approach can, in some cases, produce images with substantially higher quality compared to the method of Beltran *et al.* (2018).

To perform quantitative optimization of an imaging technique, it is necessary to define an image quality assessment (IQA) method that quantifies relevant aspects of the image quality (Cunningham & Shaw, 1999; Barrett & Myers 2004; Michail *et al.*, 2014). Generally, IQA metrics can be classified as full-reference (FR-IQA), reduced-reference (RR-IQA) and no-reference (NR-IQA) (Chow *et al*., 2016; Breger *et al*., 2025). FR-IQA metrics require presence of a "ground truth" image, such as an image obtained with an imaging system having a delta-function-like PSF and little or no image noise. In some cases, ground truth images can either be constructed numerically or obtained experimentally, using, for example, scans at high radiation doses with detectors having high spatial resolution. When a ground truth image is not available, it may be possible to use the information extracted from the image itself to assess, for example, the amount of noise in the image and then optimize the imaging setup and the post-processing methods to reduce the noise. Examples of such RR-IQA approaches include the Fourier Shell Correlation (FSC) method for measuring the spatial resolution in images (Penczek, 2010; Koho *et al*., 2019) or methods for extraction of pure noise from images used in Deep Machine Learning / Artificial Intelligence (AI) denoising (Hendriksen *et al.*, 2020; Kaur & Dong, 2023; Shah *et al*., 2025). Finally, NR-IQA approaches rely on "objective" measurements of imaging quality characteristics, such as the incident photon fluence, spatial resolution or SNR, that are substantially independent of the contents of a particular image. In other words, these quantities are considered to be characteristics of the imaging system, rather than of a particular image (Gureyev *et al.*, 2024).

Subsequently, the imaging setup and the post-processing parameters can be manipulated with the goal of optimizing these quality characteristics or metrics. Each of the three types of quality metrics has its strengths and weaknesses. Ground truth images can arguably provide the most direct and clear assessment of imaging quality. However, such images are often not available because, for example, a patient cannot be safely scanned at a high radiation dose for comparison with low-dose exposures. RR-IQA methods have the advantage of not requiring separate ground truth images, but they have to deal with the problem of separating genuine object features from noise and artefacts. NR-IQA methods can provide approaches for maximising the information channel capacity of the imaging system (Shannon, 1949; Felgett & Linfoot, 1955; Cox & Sheppard, 1986; Mikhail *et al.*, 2014; Gureyev *et al.* 2016, 2024), but they also cannot guarantee, for example, that an increase of high-frequency components in an image corresponds to real features of the imaged sample, rather than representing spurious imaging artefacts. The last problem is particularly pertinent to PBI. Indeed, the sharpening of object edges and interfaces in PBI images as a result of Fresnel diffraction objectively increases the high spatial frequency content of collected images and improves the spatial resolution. However, it can be argued that the Fresnel diffraction fringes are a kind of artefact that does not directly reflect the internal structure or shape of the imaged sample. In this respect, the key advantage of TIE-Hom retrieval is in the fact that, when applied to PBI images, it effectively trades the "excess" spatial resolution in collected PBI images for improved SNR in the reconstructed images, and in the process removes the diffraction fringes from the images (Paganin, 2006; Gureyev *et al.*, 2017). The theoretical investigations and numerical simulations presented in this paper take into account the general considerations regarding the image quality metrics discussed above.

## 2. Optimal regularization parameter in TIE-Hom retrieval

Consider a thin object located near the optical axis $z$ immediately upstream of the "object plane" $z = 0$ (Fig.1). The object is illuminated by X-rays emitted by an extended quasi-monochromatic spatially incoherent X-ray source located at $z = -R_1$. The imaged object consists of a large uniform "bulk" slab with X-ray transmission $\exp[-B_0(\lambda)]$ and an embedded small "feature of interest" with X-ray transmission $\exp[-B_0(\lambda) - B(\mathbf{r}_\perp, \lambda)]$, where $\mathbf{r}_\perp = (x, y)$ are 2D Cartesian coordinates in planes orthogonal to the optical axis. The complex X-ray transmission function of the object, $\exp[i\varphi_0(\lambda) - B_0(\lambda)/2]\exp[i\varphi(\mathbf{r}_\perp, \lambda) - B(\mathbf{r}_\perp, \lambda)/2]$, is assumed to be monomorphous in the sense that $\varphi(\mathbf{r}_\perp, \lambda) = (1/2)\gamma(\lambda)B(\mathbf{r}_\perp, \lambda)$, where $\gamma(\lambda)$ is the same at any point $\mathbf{r}_\perp$ in the object plane.

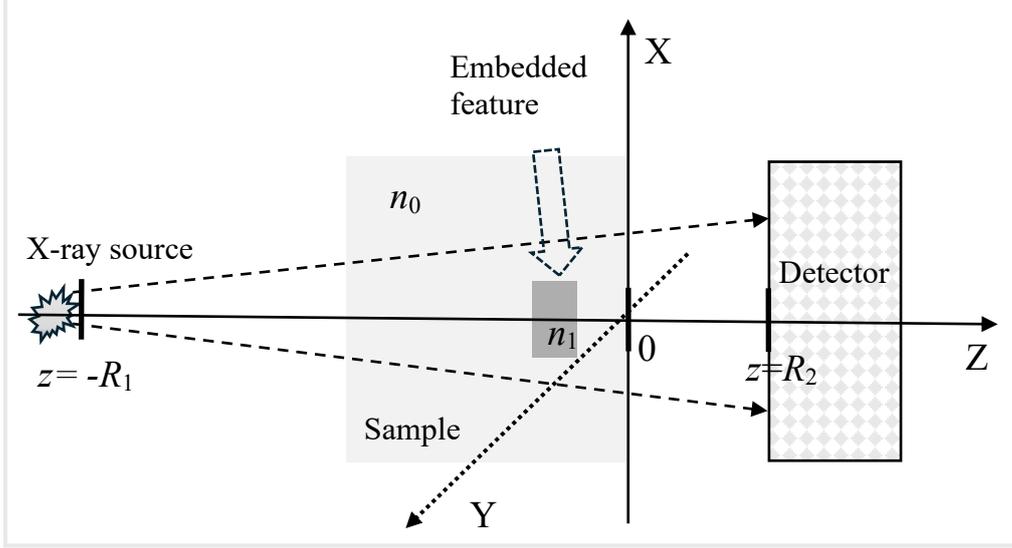

**Fig. 1.** Generic in-line imaging setup considered in the present study.

Let a 2D X-ray detector be located in the image plane $z = R_2$ (Fig.1). The detector has a quantum efficiency $\eta$ and a point-spread function (PSF) $P_{det}(\mathbf{r}_\perp)$, with unit integral $\iint P_{det}(\mathbf{r}_\perp)d\mathbf{r}_\perp = 1$ and zero means $\iint xP_{det}(\mathbf{r}_\perp)d\mathbf{r}_\perp = \iint yP_{det}(\mathbf{r}_\perp)d\mathbf{r}_\perp = 0$. This means that the PSF preserves the number of registered photons and is symmetric. The detector PSF is also assumed to be independent of the X-ray wavelength $\lambda$ for simplicity. The width of the PSF $P_{det}(\mathbf{r}_\perp)$ is defined as $\Delta_{det} = (2\pi \text{Var}[P_{det}])^{1/2}$, where $\text{Var}[P_{det}] = \iint r_\perp^2 P_{det}(\mathbf{r}_\perp)d\mathbf{r}_\perp$, $r_\perp^2 \equiv x^2 + y^2$, is the "variance" (second integral moment) of $P_{det}(\mathbf{r}_\perp)$ (Gureyev *et al.*, 2024).

The mean detected X-ray photon fluence in the object plane $z = 0$, in the vicinity of the optical axis, can be expressed as

$$\overline{I}(\mathbf{r}_\perp, 0, \lambda) = \eta \overline{I}_{id}(\mathbf{r}_\perp, 0, \lambda) * P_{det}(\mathbf{r}_\perp),$$ (1)

where $\overline{I}_{id}(\mathbf{r}_\perp,0,\lambda)=\overline{I}_{in}(\lambda)\exp[-B_0(\lambda)-B(\mathbf{r}_\perp,\lambda)]=\overline{I}_{tr}(\lambda)\exp[-B(\mathbf{r}_\perp,\lambda)]$ is the mean detected photon fluence in the object plane in the case of an ideal imaging system with a delta-function PSF and unit quantum efficiency, $I_{in}(\lambda)$ is the photon fluence of the incident beam in a sufficiently small vicinity of the optical axis where we can neglect the transverse variation of the incident intensity and $I_{tr}(\lambda)\equiv I_{in}(\lambda)\exp[-B_0(\lambda)]$ is the photon fluence of the beam transmitted through the bulk of the sample. All fluences are expressed as a number of photons per unit area.

The homogeneous Transport of Intensity equation (TIE-Hom) for the paraxial propagation of intensity of the transmitted wave (Paganin *et al*., 2002; Paganin, 2006) implies that the mean detected photon fluence in the image (detector) plane $z=R_2$ is equal to $\overline{I}(\mathbf{r}_\perp,R_2,\lambda)$, such that

$$M^2\overline{I}(M\mathbf{r}_\perp,R_2,\lambda)=\eta(1-a^2\nabla_\perp^2)\overline{I}_{id}(\mathbf{r}_\perp,0,\lambda)*P_{sys}(\mathbf{r}_\perp,M)=$$
$$\eta\overline{I}_{id}(\mathbf{r}_\perp,0,\lambda)*(1-a^2\nabla_\perp^2)P_{sys}(\mathbf{r}_\perp,M),$$
(2)

where $a^2=\gamma R'\lambda/(4\pi)$, $R'=R_2/M$, $M=(R_1+R_2)/R_1$ is the magnification and $P_{sys}(\mathbf{r}_\perp,M)=P_{det}(\mathbf{r}_\perp/M)*P_{src}(\mathbf{r}_\perp(M-1)/M)$ is the PSF of the imaging system (see e.g. (Gureyev *et al*., 2008)), where $P_{src}(\mathbf{r}_\perp)$ is the source intensity distribution, which has a unit integral, zero mean and width equal to $\Delta_{src}=(2\pi\mathrm{Var}[P_{src}])^{1/2}$. It is straightforward to verify that $P_{sys}(\mathbf{r}_\perp,M)$ also has a unit integral, zero mean and width equal to $\Delta_{sys}=[(M-1)^2M^{-2}\Delta_{src}^2+M^{-2}\Delta_{det}^2]^{1/2}$. As an example, consider the case where both the $P_{det}(\mathbf{r}_\perp)$ and $P_{src}(\mathbf{r}_\perp)$ are circular Gaussians of the form $G(\mathbf{r}_\perp,\sigma)=(2\pi\sigma^2)^{-1}\exp[-r_\perp^2/(2\sigma^2)]$. In that case, $P_{sys}(\mathbf{r}_\perp,M)$ is also a Gaussian,

$P_{sys}(\mathbf{r}_\perp,M)=[2\pi\sigma_{sys}^2(M)]^{-1}\exp\{-r_\perp^2/[2\sigma_{sys}^2(M)]\}$, with $\sigma_{sys}^2(M)=(M-1)^2M^{-2}\sigma_{src}^2+M^{-2}\sigma_{det}^2$ and the width equal to $\Delta_{sys}=(2\pi\mathrm{Var}[P_{sys}])^{1/2}=(2\pi2\sigma_{sys}^2)^{1/2}=2\pi^{1/2}\sigma_{sys}$.

Note that while eq.(2) holds for the mean values of detected fluences, it generally does not hold for the stochastic detected fluences themselves (i.e. for the detected images). In particular, the image noise in the detector plane, $M^2I(M\mathbf{r}_\perp,R_2,\lambda)-M^2\overline{I}(M\mathbf{r}_\perp,R_2,\lambda)$, is usually not equal to the result of application of the TIE-Hom operator $(1 - a^2 \nabla_\perp^2)$ to the corresponding image noise in the object plane, $\eta[I_{id}(\mathbf{r}_\perp, 0, \lambda) - \overline{I}_{id}(\mathbf{r}_\perp, 0, \lambda)] * P_{sys}(\mathbf{r}_\perp, M)$ (Gureyev *et al.*, 2024).

According to eq.(2), the PSF affecting the image in the detector plane is $(1 - a^2 \nabla_\perp^2) P_{sys}(\mathbf{r}_\perp, M)$. This function can be represented as a convolution $P_{sys}(\mathbf{r}_\perp, M) * (1 - a^2 \nabla_\perp^2) \delta_D(\mathbf{r}_\perp)$, where $\delta_D(\mathbf{r}_\perp)$ is the 2D Dirac delta-function. As variances add in a convolution operation and $\mathrm{Var}[(1 - a^2 \nabla_\perp^2) \delta_D(\mathbf{r}_\perp)] = -4a^2$ (which can be verified by direct calculations, (Gureyev *et al.*, 2024)), we can conclude that the width of the system PSF in the object plane is equal to $\mathrm{Var}[P_{sys}] - 4a^2$. This result indicates that the spatial resolution becomes finer after coherent free-space propagation (Gureyev *et al.*, 2024).

A generalized TIE-Hom retrieval operator with a regularization parameter $b$, namely $(1 - b^2 \nabla_\perp^2)^{-1}$, can be applied to the detected fluence in the image plane to produce a "reconstructed" fluence, $I_{rec}(\mathbf{r}_\perp, 0)$, in the object plane. For the mean fluences, this can be written as

$$
\begin{aligned}
\overline{I}_{rec}(\mathbf{r}_\perp, 0, \lambda) &= (1 - b^2 \nabla_\perp^2)^{-1} M^2 \overline{I}(M\mathbf{r}_\perp, R_2, \lambda) = \\
&= \eta \overline{I}_{id}(\mathbf{r}_\perp, 0, \lambda) * (1 - b^2 \nabla_\perp^2)^{-1} (1 - a^2 \nabla_\perp^2) P_{sys}(\mathbf{r}_\perp, M) = \\
&= \eta \overline{I}_{id}(\mathbf{r}_\perp, 0, \lambda) * P_{sys}(\mathbf{r}_\perp, M) * (1 - a^2 \nabla_\perp^2) \delta_D(\mathbf{r}_\perp) * (1 - b^2 \nabla_\perp^2)^{-1} \delta_D(\mathbf{r}_\perp).
\end{aligned}
\tag{3}
$$

It is easy to verify that $\mathrm{Var}[(1 - b^2 \nabla_\perp^2)^{-1} \delta_D(\mathbf{r}_\perp)] = 4b^2$. Therefore, the condition for minimising the spatial variance (i.e. for optimising the spatial resolution) of the reconstructed image distribution in the last line of eq.(3) is $\mathrm{Var}[P_{sys}] - 4a^2 + 4b^2 = 0$. Substituting $\mathrm{Var}[P_{sys}] = \Delta_{sys}^2 / (2\pi)$, we obtain

$$
b^2 = a^2 - \Delta_{sys}^2 / (8\pi).
\tag{4}
$$

If we introduce the notation $b^2 = \gamma' R' \lambda / (4\pi)$ by analogy with $a^2 = \gamma R' \lambda / (4\pi)$, then

$$
\gamma' = \gamma - \Delta_{sys}^2 / (2R'\lambda) = \gamma - N_{F,sys} / 2,
\tag{5}
$$

where $N_{F,sys} = \Delta_{sys}^2 / (R'\lambda)$ is the Fresnel number (Paganin, 2006) associated with the width of the system PSF. In order to optimize the spatial resolution in the reconstructed images, the TIE-Hom retrieval should be performed with the reduced regularization parameter defined according to eq.(5). The physical reason for the reduction of the optimal value of the regularization parameter, in comparison with that in the standard version of Paganin's method, is in the fact that the convolution of a propagated image with the system PSF is equivalent to a partial "phase retrieval". Therefore, the optimal TIE-Hom retrieval of the convolved image requires "less phase retrieval", in the sense that it requires a smaller value of the regularization parameter (Beltran *et al.*, 2018). Note that the formula corresponding to the plane-beam case of eq.(5) in (Beltran *et al.*, 2018) contains an extra factor of four compared to eq.(5), incorrectly stating that the optimal regularization parameter $\gamma'$ is equal to $\gamma - 2\Delta_{sys}^2 / (R_2\lambda)$.

Consider a definition of spatial resolution, Res, that tends to be close to the spatial resolution measured in experiments, and which is related to the width $\Delta$ as $\text{Res} \equiv \Delta / \pi^{1/2}$ (Gureyev *et al.*, 2024). In the case of a Gaussian PSF, $\text{Res}_{sys} = 2\sigma_{sys}$. As a practical example, in the case of the Xineos detector used at the Imaging and Medical Beamline (IMBL) of the Australian Synchrotron, we have $\text{Res}_{det} \cong 150\,\mu\text{m}$ (Arhatari *et al.*, 2021). Therefore, using this detector in PBI of mastectomy samples at E = 32 keV, $\gamma = 869$ (for glandular tissue embedded in adipose tissue at $E = 32$ keV), and taking into account IMBL's horizontal X-ray source size of $\Delta_{src,x} \cong \pi^{1/2} \times 800\,\mu\text{m} \cong 1418\,\mu\text{m}$, the source-to-sample distance $R_1 = 138$ m and the sample-to-detector distance $R_2 = 6$ m, we obtain $M \cong 1.0435$ and

$$N_{F,sys} = \frac{\Delta_{sys}^2(M)}{R'\lambda} = \frac{(M-1)^2\Delta_{src}^2 + \Delta_{det}^2}{MR_2\lambda} = \frac{(0.0435)^2 \times (1418\,\mu\text{m})^2 + (266\,\mu\text{m})^2}{1.0435 \times 6\,\text{m} \times 0.3875\,\text{Å}} \cong 307.$$ This

leads to the modified regularization parameter $\gamma' = \gamma - N_{F,sys} / 2 \cong 869 - (307/2) \cong 715$ that should be used in Paganin's method with this type of imaging.

Instead of optimizing the regularization parameter in TIE-Hom retrieval, eq.(3), in accordance with eqs.(4) and (5), one can instead optimize the propagation distance $R'$ in eq.(2) in order to make the Fresnel diffraction maximally counter-act the blurring by the system's PSF in the collected defocused images. This approach corresponds to "hardware" minimization of the width of the function $(1 - a^2 \nabla_\perp^2) P_{sys}(\mathbf{r}_\perp, M)$ in eq.(2), which was previously called "deblur by defocus" (Gureyev *et al.*, 2004). Similarly to eq.(4), the optimal condition in the "deblur by defocus" method can be expressed as $4a^2 - \text{Var}[P_{sys}] = 0$, or

$$R'_{opt} = \Delta_{sys}^2 / (2\gamma\lambda) \text{ or } N_{F,sys} = 2\gamma. \tag{6}$$

At the effective propagation distance $R' = R'_{opt}$ satisfying eq.(6), the Fresnel fringes appearing at edges and interfaces of a monomorphous sample as a result of free-space propagation of the coherent transmitted X-ray wave, maximally reduce (counter-act) the blurring of the edges and interfaces due to the convolution with the system's PSF. In general, however, eq.(6) does not provide an explicit solution for the optimal propagation distance, because the system resolution $\Delta_{sys}^2$ on the right-hand side of the equation depends on the propagation distance $R'$. An exception is provided by the case of plane-wave illumination, where eq.(6) does deliver an explicit solution in the form $R_{2,opt} = \Delta_{det}^2 / (2\gamma\lambda)$. In the case of a fixed source-to-detector distance $R$, an explicit solution can be formulated in terms of the optimal magnification $M_{opt} = 1 + \gamma R\lambda / \Delta_{src}^2 \pm [(\gamma R\lambda)^2 / \Delta_{src}^4 - \Delta_{det}^2 / \Delta_{src}^2]^{1/2}$, which then defines the optimal propagation distance $R_{2,opt} = R(M_{opt} - 1) / M_{opt}$. Numerical demonstration of the "deblur by defocus" effect can be found in Section 4 below, while an (indirect) experimental demonstration was previously presented in (Gureyev *et al.*, 2004). Note that optimization studies of propagation distance in PBI, including the case of noisy images, can be traced all the way back to the seminal publication by M. Teague on TIE-based phase retrieval (Teague, 1983). Other relevant works—all of which employ different approaches to those considered in the present paper—include Paganin *et al.* (2004), Eastwood *et al.* (2011), Alloo *et al.* (2022), and Miqueles *et al.* (2025).

While eqs.(4)-(6) suggest methods for optimizing spatial resolution in TIE-Hom retrieved PBI images, the spatial resolution alone does not fully determine the image quality, and other characteristics, such as SNR and radiation dose typically need to be taken into consideration as well. It was suggested previously (Gureyev *et al.*, 2016, 2024) that the ratio of squared SNR in the image to the product of the incident photon fluence and an appropriate power of the spatial resolution is closely correlated with the information channel capacity of the imaging system. Specifically, the dimensionless quantity $Q_S^2 \equiv \mathrm{SNR}^2 / (\overline{I}_{in} \Delta_{sys}^n)$, which was termed the "intrinsic imaging quality characteristic", where $n$ is the dimensionality of the images ($n = 2$ for planar images, $n = 3$ for CT images), corresponds to the amount of information regarding the imaged object that the imaging system is capable of extracting on average per single incident photon (Gureyev *et al.*, 2014, 2016). Therefore, optimization of a PBI or PB-CT imaging setup can be based on maximization of the metric $Q_S$ (Gureyev *et al.*, 2025b). However, as can be seen from eq.(4), the spatial resolution of a TIE-Hom retrieved image can formally become zero or even negative, leading to a "blow-up" (divergence) of the $Q_S$ metric. In our recent studies (Gureyev *et al.*, 2024, 2025a, 2025b), we used a modified definition of the spatial resolution (Gureyev *et al.*, 2014) that can never become zero or negative in TIE-Hom retrieval, thus eliminating the last problem. However, it can be shown that, even with use of the modified definition of spatial resolution, the magnitude of $Q_S$ can become arbitrarily large after the application of certain "natural" image processing algorithms (Gureyev *et al.*, 2026). This is a consequence of the problem mentioned in the Introduction, that is the lack of built-in discriminators in NR-IQA metrics, such as $\Delta_{sys}$ or $Q_S$, allowing one to distinguish between the true object-related spatial resolution and an artificial increase of spurious high-frequency content in reconstructed images. This problem provides a motivation for using alternative FR-IQA metrics, as studied in the next section.

## 3. Full-reference optimization of TIE-Hom retrieval

Here we consider the problem of optimal TIE-Hom reconstruction of a PBI image using a FR-IQA metric, namely the normalized root mean squared difference between the filtered (i.e. reconstructed) image $(I_R * F)(\mathbf{r}_\perp, \lambda)$, where $I_R(\mathbf{r}_\perp, \lambda) \equiv \eta^{-1} M^2 I(M\mathbf{r}_\perp, R_2, \lambda)$, and a ground truth image that would be collected in the object plane with a detector having a delta-function PSF and no image noise. According to eq.(1), such a ground truth image is equal to $\overline{I}_{id}(\mathbf{r}_\perp, 0, \lambda) = \overline{I}_{tr}(\lambda) \exp[-B(\mathbf{r}_\perp, \lambda)]$, hence the error metric can be defined as

$$RMSE[I_R * F, \overline{I}_{id}] \equiv \| I_R * F - \overline{I}_{id} \|_2 / \| \overline{I}_{id} \|_2, \tag{7}$$

where $\| f \|_p \equiv \left( \iint | f(\mathbf{r}_\perp) |^p \, d\mathbf{r}_\perp \right)^{1/p}$ and RMSE stands for "relative root mean squared error". Here $F(\mathbf{r}_\perp)$ is a "reconstruction kernel" about which we only assume for now that it has a unit integral, i.e. that convolution with $F(\mathbf{r}_\perp)$ preserves the mean number of photons in the image.

Let $\Omega$ be the minimal circle in the object plane, with the centre at the origin of coordinates, which contains all the pixels where either the reconstructed or the ground truth image is not negligibly small in magnitude. Let $\Delta I_R(\mathbf{r}_\perp, \lambda) \equiv I_R(\mathbf{r}_\perp, \lambda) - \overline{I}_R(\mathbf{r}_\perp, \lambda)$ denote the noise term in the detected image. Then $I_R(\mathbf{r}_\perp, \lambda) = \overline{I}_R(\mathbf{r}_\perp, \lambda) + \Delta I_R(\mathbf{r}_\perp, \lambda)$ and

$$MSE[I_R * F, \overline{I}_{id}] \equiv RMSE^2[I_R * F, \overline{I}_{id}] =$$
$$\| \overline{I}_{id} \|_2^{-2} \iint_\Omega [(\overline{I}_R * F)(\mathbf{r}_\perp, \lambda) + (\Delta I_R * F)(\mathbf{r}_\perp, \lambda) - \overline{I}_{id}(\mathbf{r}_\perp, 0, \lambda)]^2 d\mathbf{r}_\perp \cong$$
$$\| \overline{I}_{id} \|_2^{-2} \iint_\Omega [(\overline{I}_R * F)(\mathbf{r}_\perp, \lambda) - \overline{I}_{id}(\mathbf{r}_\perp, 0, \lambda)]^2 d\mathbf{r}_\perp + \| \overline{I}_{id} \|_2^{-2} \iint_\Omega [(\Delta I_R * F)(\mathbf{r}_\perp, \lambda)]^2 d\mathbf{r}_\perp,$$

where we assumed that when the filtered noise distribution $(\Delta I_R * F)(\mathbf{r}_\perp, \lambda)$ is modulated by the deterministic function $(\overline{I}_R * F)(\mathbf{r}_\perp, \lambda) - \overline{I}_{id}(\mathbf{r}_\perp, 0, \lambda)$, its integral is still approximately equal to zero, i.e. $\iint_\Omega [(\overline{I}_R * F)(\mathbf{r}_\perp, \lambda) - \overline{I}_{id}(\mathbf{r}_\perp, 0, \lambda)](\Delta I_R * F)(\mathbf{r}_\perp, \lambda) d\mathbf{r}_\perp \cong 0$ (see a relevant proof in Hendriksen *et al*., 2020). If X-ray absorption in the object feature is weak, we can make the approximation $\| \overline{I}_{id}(\mathbf{r}_\perp, 0, \lambda) \|_2^2 = \overline{I}_{tr}^2(\lambda) \iint_\Omega \exp[-2B(\mathbf{r}_\perp, \lambda)] d\mathbf{r}_\perp \cong \overline{I}_{tr}^2(\lambda) |\Omega|$. Since the propagation contrast is always weak under the conditions of the TIE approximation (Paganin, 2006), we can approximate the noise term by ignoring the components corresponding to the operator $-a^2 \nabla_\perp^2$ in eq.(2):

$$\Delta I_R(\mathbf{r}_\perp, \lambda) = \eta^{-1} M^2[I(M\mathbf{r}_\perp, R_2, \lambda) - \overline{I}(M\mathbf{r}_\perp, R_2, \lambda)] \cong [\eta^{-1}I_{id,\eta}(\mathbf{r}_\perp, 0, \lambda) - \overline{I}_{id}(\mathbf{r}_\perp, 0, \lambda)] * P_{sys}(\mathbf{r}_\perp, M),$$

where $I_{id,\eta}(\mathbf{r}_\perp, 0, \lambda) = I_{tr,\eta}(\lambda) \exp[-B(\mathbf{r}_\perp, \lambda)]$ is the detected photon fluence in the object plane in the case of an imaging system with a delta-function PSF and quantum efficiency equal to $\eta$. Further, we assume that the X-ray absorption in the feature of interest is sufficiently weak, so that we can neglect the effect of intensity variation in the image on the image noise and make the approximation $\Delta I_R(\mathbf{r}_\perp, \lambda) \cong [\eta^{-1}I_{tr,\eta}(\lambda) - \overline{I}_{tr}(\lambda)] * P_{det}(\mathbf{r}_\perp / M)$. Note that we have replaced $P_{sys}(\mathbf{r}_\perp, M)$ by $P_{det}(\mathbf{r}_\perp / M)$ in the last expression because, in the case of uniform illumination produced by realistic (non-laser) X-ray sources considered here, the spatially stationary image noise is affected only by the PSF of the detector, and is not affected by the intensity distribution of the X-ray source (Gureyev *et al.*, 2017).

Next, we apply Parseval's theorem to the last expression for $MSE[I_R * F, \overline{I}_{id}]$:

$$MSE[I_R * F, \overline{I}_{id}] \cong$$
$$|\Omega|^{-1} \iint |\mathbf{F}\{\exp[-B(\mathbf{r}_\perp, \lambda)]\}(\boldsymbol{\rho}_\perp)|^2 (1 + 4\pi^2 a^2 \rho_\perp^2) \hat{P}_{sys}(\boldsymbol{\rho}_\perp, M) \hat{F}(\boldsymbol{\rho}_\perp) - 1|^2 \, d\boldsymbol{\rho}_\perp + \qquad (8)$$
$$\overline{I}_{tr}^{-2}(\lambda) W(\lambda) \iint |\hat{P}_{det}(\boldsymbol{\rho}_\perp, M)|^2 |\hat{F}(\boldsymbol{\rho}_\perp)|^2 \, d\boldsymbol{\rho}_\perp,$$

where $W(\lambda)$ is the power spectral density of the noise term, $[\eta^{-1}I_{tr,\eta}(\lambda) - \overline{I}_{tr}(\lambda)]$, and $\hat{P}_{det}(\boldsymbol{\rho}_\perp, M) \equiv M^2 \hat{P}_{det}(M\boldsymbol{\rho}_\perp)$. In the first integral on the right-hand side of eq.(8), we can use the approximation $\mathbf{F}\{\exp[-B(\mathbf{r}_\perp, \lambda)]\}(\boldsymbol{\rho}_\perp) \cong \delta(\boldsymbol{\rho}_\perp) - \hat{B}(\boldsymbol{\rho}_\perp, \lambda)$ and then note that the terms with $\delta(\boldsymbol{\rho}_\perp)$ do not contribute to the integral, because $\hat{P}_{sys}(0, M)\hat{F}(0) - 1 = 0$. This reflects the fact that a constant function is invariant with respect to convolution with any PSF having unit integral, and hence for such flat objects the mean reconstructed image always coincides with the ground truth image. In the second integral in eq.(8), assuming spatially stationary Poisson statistics of detected photon fluences, we can use the fact that the power spectral density of spatially uncorrelated Poisson-distributed white noise is equal to the mean detected fluence, i.e. $W(\lambda) = \eta^{-1}\overline{I}_{tr}(\lambda)$ (Barrett & Myers, 2004). This allows us to re-write eq.(8) as

$$MSE[I_R * F, \overline{I}_{id}] \cong |\Omega|^{-1} \iint |\hat{B}(\boldsymbol{\rho}_\perp, \lambda)|^2 (1 + 4\pi^2 a^2 \rho_\perp^2) \hat{P}_{sys}(\boldsymbol{\rho}_\perp, M) \hat{F}(\boldsymbol{\rho}_\perp) - 1|^2 \, d\boldsymbol{\rho}_\perp +$$
$$\eta^{-1}\overline{I}_{tr}^{-1}(\lambda) \iint |\hat{P}_{det}(\boldsymbol{\rho}_\perp, M)|^2 |\hat{F}(\boldsymbol{\rho}_\perp)|^2 \, d\boldsymbol{\rho}_\perp. \qquad (9)$$

The first (deterministic) additive term on the right-hand side of eq.(9) depends on the difference between the Fourier transforms of the reconstructed PSF, $(1 + 4\pi^2 a^2 \rho_\perp^2) \hat{P}_{sys}(\boldsymbol{\rho}_\perp, M) \hat{F}(\boldsymbol{\rho}_\perp)$, and the delta-function PSF of the ground truth imaging system: the error is proportional to this difference. This term is also affected by the Fourier transform of the absorption contrast $1 - \exp[-B(\mathbf{r}_\perp, \lambda)] \cong B(\mathbf{r}_\perp, \lambda)$ produced by the feature of interest: if this contrast is weak or slowly varying, the first error term can be small even if the reconstructed PSF is very different from the delta-function. This error term is also small when the contrast is slowly varying because $(1 + 4\pi^2 a^2 \times 0) \hat{P}_{sys}(0, M) \hat{F}(0) - 1 = 0$. The second (stochastic) additive error term on the right-hand side of eq.(9) is proportional to the variance of the noise in the collected PBI image, which is inversely proportional to the transmitted X-ray fluence, the quantum efficiency of the detector and the effective "pixel size" in the reconstructed image. In turn, the "reconstructed pixel size" is determined by the convolution of the detector PSF and the reconstruction kernel $F(\mathbf{r}_\perp)$: the broader the convolution, the larger the effective pixel size and the smaller the resultant noise variance (due to increased spatial correlation of the photon fluence). The last fact implies a clear trade-off between the first and second error terms in eq.(9): when the Fourier transform of the reconstruction kernel, $\hat{F}(\boldsymbol{\rho}_\perp)$, is close to the inverse of the Fourier transform of the forward-imaging PSF, $(1+4\pi^2 a^2 \rho_\perp^2)\hat{P}_{sys}(\boldsymbol{\rho}_\perp, M)\hat{F}(\boldsymbol{\rho}_\perp)$, the first error term becomes small, but the second error term may become large. On the other hand, when $\hat{F}(\boldsymbol{\rho}_\perp)$ is close to a delta-function, the second error becomes smaller, but the first error term can become large.

When the noise term, i.e. the second integral on the right-hand side of eq.(9), is negligibly small, the optimal reconstruction kernel is given by $\hat{F}(\boldsymbol{\rho}_\perp) = \hat{P}_{sys}^{-1}(\boldsymbol{\rho}_\perp, M)(1+4\pi^2 a^2 \rho_\perp^2)^{-1}$. This corresponds to the conventional TIE-Hom retrieval in combination with deconvolution of the system PSF. Such a reconstruction kernel makes the first integral in the right-hand side of eq.(9) equal to zero. However, in the presence of non-negligible image noise, a non-regularized deconvolution of the system PSF can amplify noise, making the second integral on the right-hand side of eq.(9) large in magnitude. This suggests the need to consider a regularized version of the reconstruction filter, such as, for example, Tikhonov's regularization (Tikhonov & Arsenin, 1979) or the Moore-Penrose pseudo-inverse (Barrett & Myers, 2004). Accordingly, in the present case one can use the regularized reconstruction kernel

$$\hat{F}_\varepsilon(\boldsymbol{\rho}_\perp) = \frac{(1+\varepsilon)\hat{P}_{sys}^\dagger(\boldsymbol{\rho}_\perp, M)}{(1+4\pi^2 a^2 \rho_\perp^2)[|\hat{P}_{sys}(\boldsymbol{\rho}_\perp, M)|^2 + \varepsilon]}, \qquad (10)$$

where "$\dagger$" denotes complex conjugation and $\varepsilon > 0$ is a (constant) regularization parameter. Note that we included the term $(1 + \varepsilon)$ in the numerator in eq.(10) in order to preserve the normalization $\hat{F}_\varepsilon(0) = 1$ for any $\varepsilon$.

Substituting $\hat{F}(\boldsymbol{\rho}_{\perp}) = \hat{F}_{\varepsilon}(\boldsymbol{\rho}_{\perp})$ into eq.(9), we obtain:

$$MSE[I_R * F, \overline{I}_{id}] \cong |\Omega|^{-1} \varepsilon^2 \iint \frac{|\hat{B}(\boldsymbol{\rho}_{\perp}, \lambda)|^2 [|\hat{P}_{sys}(\boldsymbol{\rho}_{\perp}, M)|^2 - 1]^2}{[|\hat{P}_{sys}(\boldsymbol{\rho}_{\perp}, M)|^2 + \varepsilon]^2} d\boldsymbol{\rho}_{\perp} +$$

$$\eta^{-1} \overline{I}_{tr}^{-1}(\lambda)(1+\varepsilon)^2 \iint \frac{|\hat{P}_{det}(\boldsymbol{\rho}_{\perp}, M)|^2 |\hat{P}_{sys}(\boldsymbol{\rho}_{\perp}, M)|^2}{(1 + 4\pi^2 a^2 \rho_{\perp}^2)^2 [|\hat{P}_{sys}(\boldsymbol{\rho}_{\perp}, M)|^2 + \varepsilon]^2} d\boldsymbol{\rho}_{\perp}. \tag{11}$$

The first integral on the right-hand side of eq.(11) is determined by the absorption contrast, $B(\mathbf{r}_{\perp}, \lambda)$, and by the difference between the system PSF and the Dirac delta function. As shown in Appendix A, the second integral in eq.(11) is inversely proportional to the squared SNR in the TIE-Hom reconstructed image, $SNR_{TIE}^2 = G_2^2 SNR_0^2$, where $SNR_0^2 = \eta \overline{I}_{tr}(\lambda) \Delta_{det}^2$ is the squared SNR in in the contact image and $G_2 \equiv (\gamma / N_{F,pix})^{1/2} = (\gamma R' \lambda / \Delta_{pix}^2)^{1/2}$ is the PBI "gain factor" (Nesterets & Gureyev, 2014; Gureyev $et\ al.$, 2024). When $\varepsilon \to 0$, the first additive term in eq.(11) typically tends to zero, while the second term can be large if $\hat{P}_{sys}(\boldsymbol{\rho}_{\perp}, M)$ has zero or near-zero values at some points.

It is useful to find at least an order-of-magnitude estimate for the optimal value of the regularization parameter $\varepsilon$ in eq.(11) analytically, so that it could then be numerically refined for specific experimental conditions, as needed. A relevant result is derived in Appendix A:

$$\varepsilon_{opt} = \frac{1}{C_B^2 SNR_{TIE}^2 (1 - \Delta_{sys,pix}^{-2}) - 1}, \tag{12}$$

where $C_B^2 \equiv |\Omega|^{-1} \iint |\hat{B}(\boldsymbol{\rho}_{\perp}, \lambda)|^2 d\boldsymbol{\rho}_{\perp} = |\Omega|^{-1} \iint B^2(\mathbf{r}_{\perp}, \lambda) d\mathbf{r}_{\perp}$ has the meaning of average squared absorption contrast produced by the object feature and $\Delta_{sys,pix} \equiv \Delta_{sys} / \Delta_{pix}$ is the ratio of the widths of the system PSF and the detector pixel. The regularization parameter from eq.(12) can be used in the reconstruction of an object-plane image in the form of the convolution $(I_R * F_{\varepsilon_{opt}})(\mathbf{r}_{\perp}, \lambda)$ of the PBI image $I_R(\mathbf{r}_{\perp}, \lambda)$ collected at the image plane $z = R_2$, with the regularized reconstruction kernel $F_{\varepsilon}(\mathbf{r}_{\perp})$ defined by eq.(10). In the next section, we will test this method for PBI image reconstruction with the help of a numerically simulated imaging setup and a simulated weakly-absorbing sample, and compare the results with the

corresponding reconstructions obtained using the conventional TIE-Hom reconstruction and its optimized versions defined by eqs.(3)-(6).

## 4. Numerical tests

Here we present the outcomes of tests of the theoretical results obtained in the previous two sections regarding the optimized variants of TIE-Hom retrieval. We used a simulated imaging setup roughly corresponding to Fig.1 in the case of a monochromatic plane incident X-ray wave with wavelength $\lambda = 1$ Å. The source-to-sample distance in the plane-wave geometry was formally infinite, $R_1 = \infty$, and the sample-to-detector distance was $R_2 = 1.0$ m. A numerically simulated homogeneous sample had a transmission function with the approximate range of transmission values (0.9, 1.0) and with $\gamma = 100$ at the chosen wavelength $\lambda = 1$ Å. The transmitted intensity in the object plane is shown in Fig.2(a), which represents the ground truth image corresponding to a delta-function PSF and no noise. The detector was assumed to have perfect quantum efficiency, $\eta = 1$, square pixels with size $10 \times 10$ μm$^2$ and a circular Gaussian PSF with $\sigma_{det} = 20$ μm, i.e. with spatial resolution Res$_{det} = 40$ μm (equal to four pixels). The registered "flat-field" photon fluence (without the object in the beam) was assumed to be Poisson distributed with a mean value of 100 photons per detector pixel. A simulated noisy image in the object plane, convolved with the detector PSF, is shown in Fig.2(b) in the form of the photon fluence normalized by the mean number of detected photons per pixel in the flat-field image. We also numerically simulated the corresponding PBI image in the plane $R_2 = 1.0$ m with the same detector PSF and photon fluence (Fig.2(c)). The simulation of free-space propagation was performed by computing Fresnel diffraction integrals on a square numerical grid with a period of 2.5 μm, followed by downsampling to the pixel size of 10 μm, simulation of pseudo-random Poisson noise with mean equal to 10% of pixel values and, finally, the application of a Gaussian low-pass filter with Res$_{det} = 40$ μm. By comparing the images in Fig.2(b) and Fig.2(c), it is easy to see that the image contrast and contrast-to-noise ratio have increased after the coherent free-space propagation. The characteristic white-black Fresnel diffraction fringes are visible at the interfaces between different components of the propagated image that correspond to areas with different X-ray attenuation coefficients in the simulated object. These fringes largely disappear in the next simulated image shown in Fig.2(d), which corresponded to the plane $R_{2,opt} \cong 25.1$ cm. The optimal propagation distance $R_{2,opt} = \pi \mathrm{Res}_{det}^2 / (2\gamma\lambda)$ was calculated here in accordance with eq.(6) for the case of plane-wave illumination. At the same time, the edges and interfaces in Fig.2(d) are still noticeably sharper than in Fig.2(b). This demonstrates the "deblur by defocus" effect (Gureyev *et al.*, 2004), involving the optimal cancellation of the PSF-induced blurring by the propagation-induced Fresnel diffraction fringes at the optimal propagation distance.

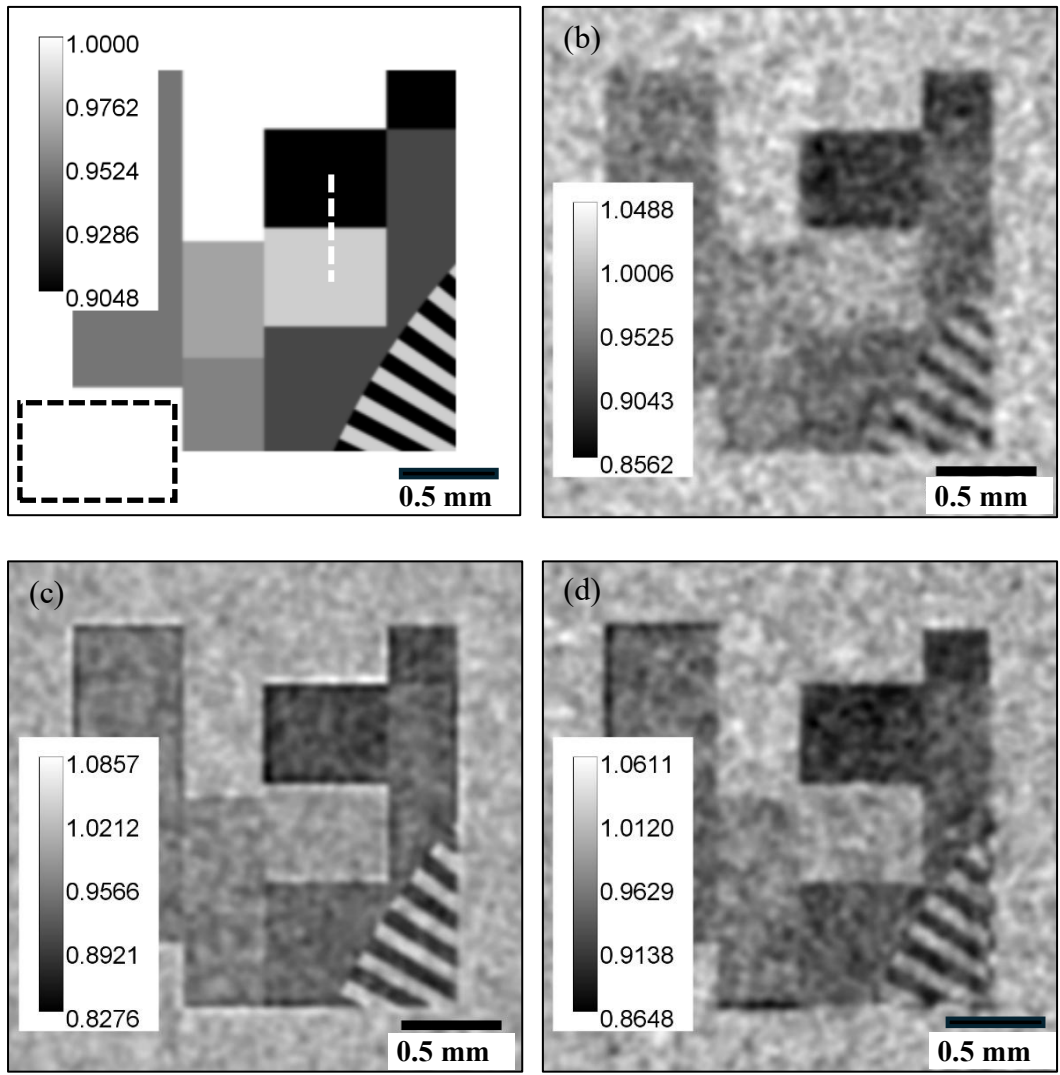

**Fig. 2.** (a) The original ground truth image in the plane $R_2 = 0$ corresponding to delta-function PSF and no noise. The white dashed line indicates the interval across which the spatial resolution was measured in the images. The black dashed rectangle indicates the area in which the SNR was measured in the images. (b) Normalized detected image in the plane $R_2 = 0$ with a 4-pixel wide Gaussian PSF and 10% Poisson noise. (c) Normalized detected image in the plane $R_2 = 1$ m with a 4-pixel wide Gaussian PSF and 10% Poisson noise. (d) Normalized detected image in the plane $R_{2,opt} \cong 25.1$ cm with a 4-pixel wide Gaussian PSF and 10% Poisson noise. The grey scale level calibration in each image is linear and corresponds to the full image histogram.

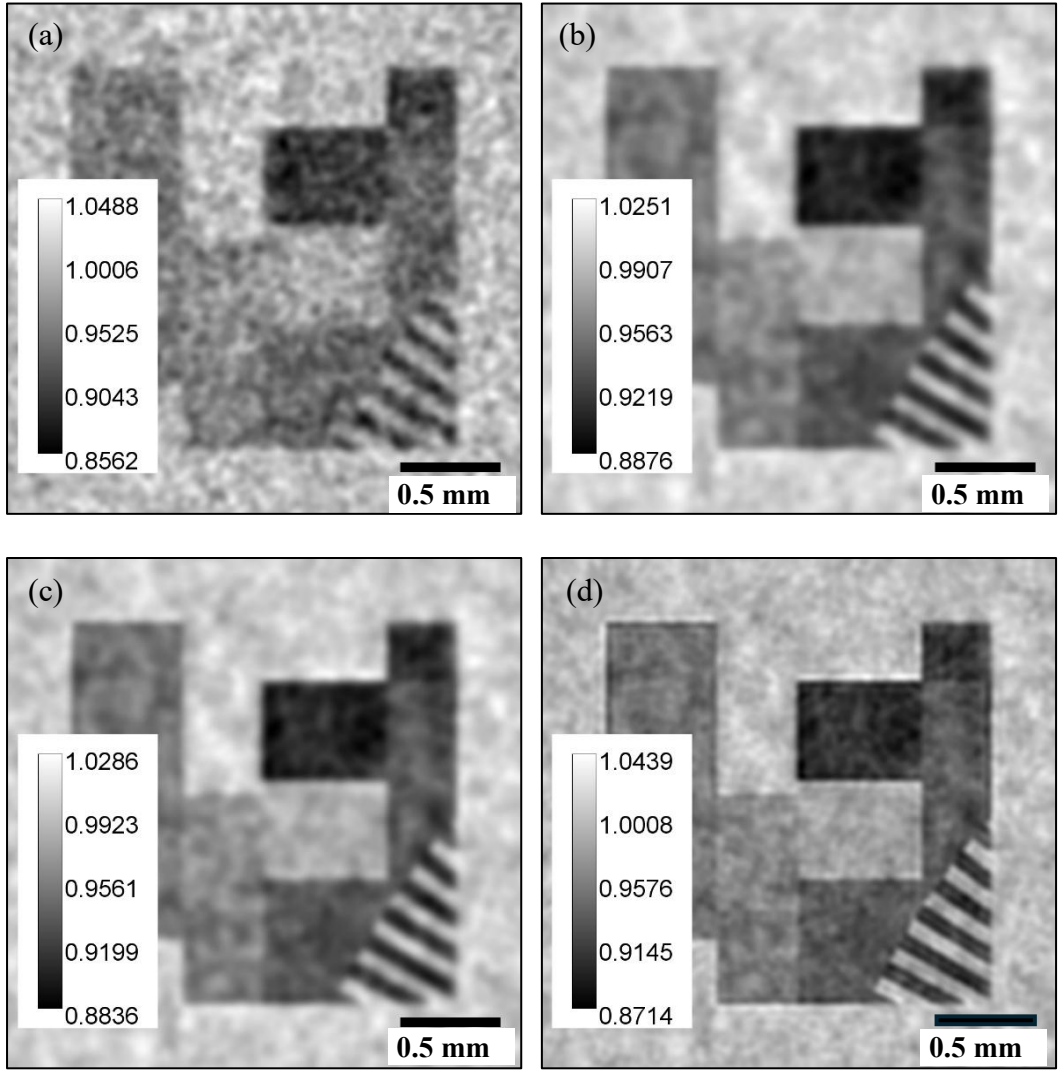

**Fig. 3.** (a) Normalized detected image in the plane $R_2 = 0$ with a 4-pixel wide Gaussian PSF and 10% Poisson noise. (b) TIE-Hom reconstruction from image in Fig.2(c) with $\gamma = 100$. (c) TIE-Hom reconstruction from image in Fig.2(c) with $\gamma' = 75$. (d) TIE-Hom reconstruction from image in Fig.2(c) with $\gamma = 100$ and Tikhonov deconvolution of the system PSF with $\varepsilon_{opt} = 0.033$. The grey scale level calibration in each image is linear and corresponds to the full image histogram.

The next several images presented in Fig.3 illustrate the effect of TIE-Hom retrieval. The image in Fig.3(a) reproduces the same detected image in the object plane $R_2 = 0$ with a 4-pixel wide Gaussian PSF and 10% Poisson noise as in Fig.2(b), which is included here for the ease of comparison with the related TIE-Hom reconstructed images. Figure 3(b) shows the result of application of Paganin's method in its standard form, i.e. the result of TIE-Hom retrieval with $\gamma = 100$ from the PBI image collected in the image plane $R_2 = 1.0$ m with the same detector PSF and photon fluence as in the object plane. It is easy to see that the TIE-Hom retrieved image in Fig.3(b) has similar spatial resolution and much higher SNR compared to the "contact" image in Fig.3(a). This result illustrates the known advantage of increased image quality obtainable in TIE-Hom imaging using Paganin's retrieval method (Paganin, 2006). Figure 3(c) presents a TIE-Hom image reconstructed from the same propagated image (shown in Fig.2(c)), but with a reduced regularization parameter $b^2 = \gamma' R' \lambda / (4\pi)$, with the optimal value of $\gamma' = 75$ that was calculated according to eq.(5). By comparing the images in Fig.3(b) and Fig.3(c), it is possible to notice that the edges and interfaces in Fig.3(c) are slightly sharper than in Fig.3(b), while the SNR is slightly reduced. These observations are confirmed by the results of quantitative measurements of SNR and spatial resolution presented in Table 1 below.

Improvement of image quality was much more pronounced in the calculated TIE-Hom image $I_{0,rec}(\mathbf{r}_\perp, \lambda) = (I_R * F_{\varepsilon_{opt}})(\mathbf{r}_\perp, \lambda)$ shown in Fig.3(d), which was obtained using the optimized reconstruction kernel $F_{\varepsilon_{opt}}(\mathbf{r}_\perp)$ structured according to eq.(10) with $\gamma = 100$ and with the Tikhonov regularization parameter calculated using eq.(12):

$\varepsilon_{opt} = \dfrac{1}{C_B^2 SNR^2 (1 - \Delta_{det,pix}^{-2}) - 1} \cong \dfrac{1}{0.0017 \times 140^2 \times (15/16) - 1} \cong 0.033$ . Here we used the average squared contrast value $C_B^2 \equiv |\Omega|^{-1} \iint B^2(\mathbf{r}_\perp, \lambda) d\mathbf{r}_\perp \cong 0.0017$ measured in the image in Fig.2(a), the value of SNR = 140 measured in the image in Fig.3(b) (see also Table 1 below) and the fact that in the present numerical example we had $\Delta_{sys} / \Delta_{pix} = \Delta_{det} / \Delta_{pix} = 4$ .

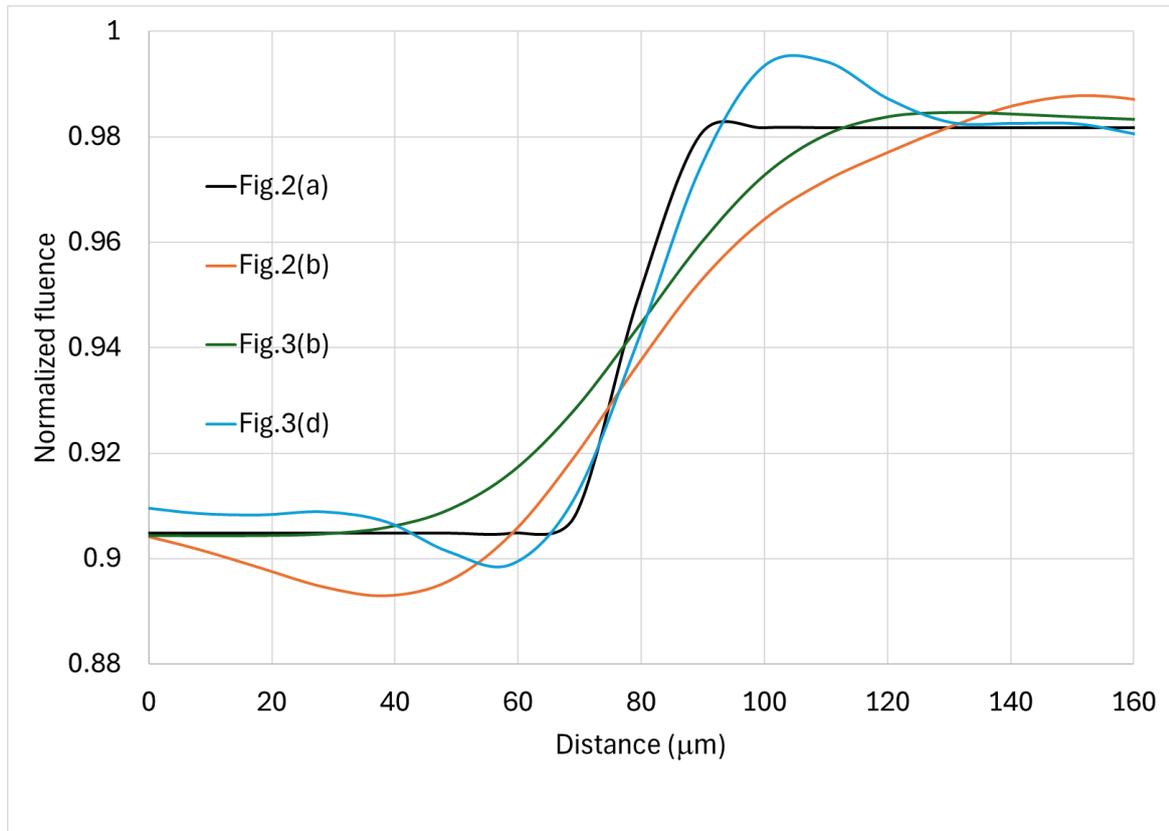

**Fig.4.** Vertical cross-sections along the interval shown by the white dashed line in Fig.2(a), averaged along the horizontal direction, for the images from Fig.2(a) (black line), Fig.2(b) (orange line), Fig.3(b) (green line) and Fig.3(d) (blue line). These profiles reveal the width of the images of the interface between two adjacent areas with different X-ray absorptions, which allows one to estimate the spatial resolution.

The relative spatial resolutions of the images shown in Figs.2(a), 2(b), 3(b) and 3(d) can also be compared in Fig.4, which presents linear cross-sections measured in these images along the interval shown by the white dashed line in Fig.2(a), averaged over the horizontal direction to reduce the visible noise in the cross-sections. Quantitative results of measurements of SNR, spatial resolution and other metrics in the images from Figs.2-3 can be found in Table 1. These results correlate well with the subjective visual assessment of these images discussed above.

**Table 1.** SNR, spatial resolution, RMSE and SSIM in the contact ($R = 0$) and TIE-Hom reconstructed images with a 4-pixel wide Gaussian PSF and 10% Poisson noise. The SNR was measured as the ratio of the mean value to the standard deviation of pixel intensities within the rectangle outlined by the black dashed line in Fig.1(a). The spatial resolution $\Delta$ was measured by evaluating the broadening of the interface, in comparison with Fig.1(a), at the location shown by the white dashed line in Fig.1(a) (see also Fig.4).

| Image | SNR | $\Delta$ ($\mu$m) | SNR/$\Delta$ ($\mu$m$^{-1}$) | RMSE | SSIM |
|---|---|---|---|---|---|
| **Fig. 2(a)** | N/A | 0 | N/A | 0 | 1.000 |
| **Fig. 2(b)** | 71.7 | 44.2 | 1.62 | 0.123 | 0.893 |
| **Fig. 2(c)** | 70.7 | 16.5 | 4.28 | 0.130 | 0.890 |
| **Fig. 2(d)** | 66.5 | 34.3 | 1.94 | 0.132 | 0.888 |
| **Fig. 3(b)** | 140 | 32.9 | 4.26 | 0.100 | 0.967 |
| **Fig. 3(c)** | 127 | 30.0 | 4.23 | 0.101 | 0.966 |
| **Fig. 3(d)** | 106 | 14.7 | 7.21 | 0.104 | 0.954 |

In summary, the data in Table 1 show that the SNR remained unchanged after forward coherent free-space propagations, while the spatial resolution improved significantly compared to the object-plane image. After TIE-Hom retrieval, the SNR significantly increased, while the spatial resolution deteriorated. Note that the ratio SNR / $\Delta$ increased after the free-space propagation (in Fig.2(c) compared to Fig.2(b)), but then remained remarkably stable after TIE-Hom retrieval with different parameters (see Fig.3(b) and Fig.3(c)). This behaviour is in agreement with the known theoretical invariance of SNR / $\Delta$ and $Q_S$ metrics in linear transformations of images (Gureyev *et al.*, 2016). The best-quality image, according to the SNR / $\Delta$ metric, was obtained with the optimized TIE-Hom retrieval using eqs.(11) and (13) (Fig.3(d)), which demonstrated a higher SNR compared to the corresponding object-plane image (Fig.2(b)), as well as a significantly improved spatial resolution. In particular, the improved spatial resolution was a direct result of numerical deconvolution with the system PSF, where the associated amplification of noise was minimized by applying the Tikhonov regularization with the optimized regularization parameter. The SNR / $\Delta$ ratio was considerably higher in Fig.3(d) than in all other images, including Figs.2(c), 3(b) and 3(c). This does not contradict the theoretical invariance of this quantity that was mentioned above, because the regularized deconvolution is a non-linear transformation. We also included the results for the Structural Similarity Index Measure (SSIM) (Wang *et al.*, 2004) in Table 1 for comparison with other metrics, since the SSIM is very popular in the image processing and analysis community.

The SSIM and RMSE metric values in Table 1 indicate that coherent free-space propagation in combination with TIE-Hom retrieval noticeably improved the similarity between the reconstructed images and the ground truth image, compared to the image collected in the object plane with the same detector resolution and the same photon flux. On the other hand, the SSIM and RMSE metrics of the image reconstructed according to eqs.(11) and (13) in Fig.3(d), were marginally worse than the conventional TIE-Hom reconstructed image, Fig.3(b), despite the image Fig.3(d) having the best SNR / Δ ratio. Further investigation revealed that, in the current numerical example, the SSIM and RMSE metrics were dominated by the image noise, while the contrast differences did not have a substantial effect, because they were only pronounced along the interfaces between regions with different X-ray absorption. The RMSE metric was also rather insensitive to changes of the regularization parameter $\varepsilon$ in eq.(11). Making the regularization parameter in the deconvolution smaller or larger than the estimated optimal value $\varepsilon_{opt} \cong 0.033$ by an order of magnitude left the value of the RMSE metric unchanged within the numerical precision of our calculations. This was despite the fact that the visual sharpness of the reconstructed images, as well as the numerical values of the SNR and spatial resolution, varied significantly with such changes of $\varepsilon_{opt}$. This outcome indicates that the RMSE and SSIM metrics may not be the best tools for evaluation and optimization of TIE-Hom reconstruction algorithms, due to the low sensitivity of these metrics to the variation of contrast, resolution and perceived visual quality of such images (because the main differences between these images appear only near the borders and interfaces).

## 5. Discussion and conclusions

We have studied some options for optimising the performance of Paganin's method for image reconstruction in PBI and PB-CT with spatially coherent X-ray illumination. We suggested that, in general, an optimization of this kind requires an image quality metric for quantitative assessment of the performance of proposed algorithms for image acquisition and processing.

We have recently conducted a detailed study (Gureyev *et al.*, 2025b) of optimization of PBI and PB-CT imaging setups on the basis of simple NR-IQA metrics, such as contrast, SNR, spatial resolution, radiation dose, and related combined metrics, including the intrinsic imaging quality (Gureyev *et al.*, 2016) and biomedical X-ray imaging quality (Gureyev *et al.*, 2025a). The present work complements the study (Gureyev *et al.*, 2025b) by analysing approaches to optimization of computer processing (reconstruction) of PBI and PB-CT images. Such approaches are aimed at improving the standard Paganin method for image retrieval (Paganin *et al.*, 2002; Paganin, 2006). We first considered an optimization of this method using the spatial resolution of the reconstructed image as a primary (NR-IQA) quality metric, where we followed the approach originally proposed in (Beltran *et al.*, 2018). We also provided explicit analytical expressions for the defocus distance at which the Fresnel diffraction fringes optimally counteract the image blurring induced by the system PSF, which corresponds to the "deblur by defocus" method (Gureyev *et al.*, 2004). We then studied an alternative optimization of Paganin's method using a FR-IQA RMSE metric, defined in eq.(7), for the distance between the reconstructed image $I_R * F$ and the ground truth image $\bar{I}_{id}$ that would be obtained in an imaging system with a delta-function PSF and no noise. We showed that this last metric inherently takes into account the contrast, SNR and spatial resolution, just as the (NR-IQA) intrinsic imaging quality metric $Q_S$ does. However, unlike $Q_S$, the RMSE metric naturally discriminates between the true spatial resolution and SNR, associated with the imaged object, and the contribution to these metrics from artefacts that can be present in a reconstructed image. This discrimination is achieved due to the use of the ground truth image in the RMSE metric. When such a ground truth image is not available, this metric obviously cannot be used, and one may have to use instead a NR-IQA metric like $Q_S$, or RR-IQA metrics, where, for example, the noise level is estimated from the image or the CT data itself (Hendriksen *et al.*, 2020; Shah *et al.*, 2025). Furthermore, the optimized TIE-Hom reconstruction $I_R * F_{\varepsilon_{opt}}$, which minimized the RMSE metric, required sufficiently detailed and accurate knowledge of the system PSF. When such information is not available, and, for example, only the width of the system PSF is known, a cruder solution based on eqs.(3)-(5) or eq.(6) can still be applied. For noisy images, the value of the optimal parameter $\gamma'$ given by eq.(5) can be increased in order to more strongly suppress the image noise in the process of TIE-Hom retrieval. However, we suggested that it is impossible to find an optimal parameter $\gamma'$ that would maximize the NR-IQA intrinsic imaging quality metric $Q_S$, or the biomedical imaging quality metric (Gureyev *et al.*, 2025), because these metrics diverge when the width of the PSF associated with the TIE-Hom retrieved image formally tends to zero. As explained above, the last problem is related to the general inability of NR-IQA metrics to distinguish between the true features of the image that correspond to the imaged sample and extraneous features, such as noise or deterministic artefacts. This represents, arguably, the main limitation of NR-IQA metrics, which can be overcome by the use of FR-IQA metrics. However, apart from the issue of availability of ground truth references for analysis, we pointed (at the end of Section 5) to another potential issue with some FR-IQA metrics, namely their low sensitivity to certain types of variations in the reconstructed images that strongly affect their SNR, spatial resolution and the overall perceived image quality (Chow *et al*., 2016; Breger *et al*., 2025). In our numerical simulations, while the SNR and spatial resolution varied significantly as functions of the image reconstruction parameters, the RMSE metric was found to be almost constant within a wide range of parameter changes.

The advantages and disadvantages of various image quality metrics ("loss functions") have to be taken into account when AI models are trained for the purpose of tasks such as image denoising or segmentation (Hendriksen *et al.*, 2021; Pakzad *et al*., 2025; Shah *et al.*, 2025). In that respect, we believe that the results of the present study can be useful in future research aimed at developing better AI methods for denoising and segmenting PBI and PB-CT images. This direction of research has shown great promise in recent years and is being actively investigated at present. Accurate and sensitive image quality metrics, which are also resilient to image artefacts, are very important in that work, as they can optimally drive the process of adjusting model parameters during iterative training cycles. On the other hand, any deficiencies in the metrics, such as, for example, a propensity to indiscriminately favour high-frequency or low-frequency content in the images, can easily divert the progress of AI model training to incorrect targets and lead to suboptimal outcomes.

## Appendix A. Derivation of optimum value of the Tikhonov regularization parameter

Here we first simplify eq.(11) by assuming that the system and the detector modulation transfer functions vary so slowly as function of transverse position that they can be approximated over the whole integration area by their values at some "point of average value", $\bar{\boldsymbol{\rho}}_\perp$. We can then take the quantities $s \equiv |\hat{P}_{sys}(\bar{\boldsymbol{\rho}}_\perp, M)|^2$ and $d \equiv |\hat{P}_{det}(\bar{\boldsymbol{\rho}}_\perp, M)|^2$ out of the integrals in eq.(11) to obtain

$$MSE[I_R * F, \bar{I}_{id}] \cong \frac{C_B^2(1-s)^2 \varepsilon^2}{(s+\varepsilon)^2} + \frac{sd}{SNR_{pix}^2 G_2^2} \frac{(1+\varepsilon)^2}{(s+\varepsilon)^2}. \tag{A1}$$

Here $C_B^2 \equiv |\Omega|^{-1} \iint |\hat{B}(\boldsymbol{\rho}_\perp, \lambda)|^2 d\boldsymbol{\rho}_\perp = |\Omega|^{-1} \iint B^2(\mathbf{r}_\perp, \lambda) d\mathbf{r}_\perp$ has the meaning of average squared absorption contrast produced by the object feature and $SNR_{pix}^2 G_2^2$ corresponds to the squared SNR in the "standard" TIE-Hom reconstructed image in the case of a detector with a PSF whose width is equal to that of a single pixel, where $SNR_{pix}^2 = \bar{n}_{pix} = \eta \bar{I}_{tr}(\lambda) \Delta_{pix}^2$ is the mean number of photons registered in one detector pixel in the object plane and $\Delta_{pix}^2 \iint (1 + 4\pi^2 a^2 \rho_\perp^2)^{-2} d\boldsymbol{\rho}_\perp = \Delta_{pix}^2 / (4\pi a^2) = \Delta_{pix}^2 / (\gamma R' \lambda) = N_{F,pix} / \gamma = 1 / G_2^2$ is the inverse of the squared 2D gain factor $G_2 \equiv (\gamma / N_{F,pix})^{1/2}$ (Nesterets & Gureyev, 2014; Gureyev *et al.*, 2024). The right-hand side of eq.(A1) can be minimized with respect to $\varepsilon$ by finding the zero points of the first derivative of this expression with respect to $\varepsilon$, which happens to have a unique solution

$$\varepsilon_{opt} = \frac{1}{C_B^2 SNR_{pix}^2 G_2^2 (1-s) d^{-1} - 1}. \tag{A2}$$

This estimated optimal value of the regularization parameter possesses the expected properties, in that it is inversely proportional to the squared image contrast and to the squared SNR in the reconstructed image.

In a typical case, the MTFs $|\hat{P}_{sys}(\boldsymbol{\rho}_\perp, M)|$ and $|\hat{P}_{det}(\boldsymbol{\rho}_\perp, M)|$ have their maximum values at the zero spatial frequency, where $|\hat{P}_{sys}(0, M)| = |\hat{P}_{det}(0, M)| = 1$ due to the normalization of the corresponding PSFs. The MTFs also tend to zero at high spatial frequencies. Therefore, we can assume that $0 < s, d < 1$. For the purpose of an order-of-magnitude estimation, we can

choose the average values $s = d = 0.5$, which leads to $\varepsilon_{opt} \sim 1/(C_B^2 SNR_{pix}^2 G_2^2 - 1)$. However, it is also possible to obtain a further approximation by relating the parameters $s$ and $d$ to the spatial resolution in the reconstructed image. Let $\hat{\Omega}$ be the minimal circle in the reciprocal (Fourier) space outside which the Fourier transforms of the reconstructed and ground truth images are both zero or negligibly small, so the integrals in eq.(11) can be restricted to $\hat{\Omega}$. In the case of discrete (pixellated) images, we can assume that $\hat{\Omega}$ is a circle with radius $\rho_{\perp,max} \equiv 1/(\sqrt{\pi}\Delta_{pix})$, where $\Delta_{pix}$ is the size of the detector pixel, and hence the area of $\hat{\Omega}$ is equal to $|\hat{\Omega}| = \pi\rho_{\perp,max}^2 = \Delta_{pix}^{-2}$. The average value of the MTF $|\hat{P}_{sys}(\boldsymbol{\rho}_\perp, M)|$ can then be approximated as $s \approx |\hat{\Omega}|^{-1} \iint_{\hat{\Omega}} |\hat{P}_{sys}(\boldsymbol{\rho}_\perp, M)|^2 \, d\boldsymbol{\rho}_\perp = \Delta_{pix}^2 \iint_{\Omega} |P_{sys}(\mathbf{r}_\perp, M)|^2 \, d\mathbf{r}_\perp = \Delta_{pix}^2 / \tilde{\Delta}_{sys}^2$, where $\tilde{\Delta}_{sys} \equiv \| P_{sys} \|_1 / \| P_{sys} \|_2$ as used in (Gureyev *et al.*, 2016; Gureyev *et al.*, 2024) and other publications, and $\| P_{sys} \|_1 = 1$ by definition. The same approach can be applied to $\hat{P}_{det}(\boldsymbol{\rho}_\perp, M)$, resulting in $d \approx \Delta_{pix}^2 / \tilde{\Delta}_{det}^2$. It has been shown (Gureyev *et al.*, 2016; 2024) that the metrics $\Delta_{sys}$ and $\tilde{\Delta}_{sys}$ produce the same or very similar estimations for the width of many functional forms of PSFs; for example, $\tilde{\Delta}_{sys} = \Delta_{sys}$ for any Gaussian PSF. Therefore, we shall assume that $s \approx \Delta_{pix}^2 / \Delta_{sys}^2$ and $d \approx \Delta_{pix}^2 / \Delta_{det}^2$. Note that these values still satisfy the inequalities $0 < s, d < 1$. Then $(1-s)d^{-1} = \Delta_{det,pix}^2 (1 - 1/\Delta_{sys,pix}^2)$, where we have introduced a special notation for spatial resolutions of the system and the detector measured in units of detector pixels: $\Delta_{sys,pix} \equiv \Delta_{sys} / \Delta_{pix}$ and $\Delta_{det,pix} \equiv \Delta_{det} / \Delta_{pix}$. Substituting this result into eq.(A2) and taking into account the fact that $G_2^2 SNR_{pix}^2 \Delta_{det,pix}^2 = G_2^2 SNR_0^2 = SNR_{TIE}^2$ is the squared SNR in the TIE-Hom reconstructed image (Nesterets & Gureyev, 2014) and $SNR_0^2 \equiv SNR_{pix}^2 \Delta_{det,pix}^2 = \eta \bar{I}_{tr}(\lambda) \Delta_{det}^2$ is the squared SNR in the contact image, we obtain eq.(12).